# A Survey on Application Layer Protocols for IoT Networks

Fatma Hmissi
*Cristal Laboratory, ENSI*
*Manouba University Campus*
Manouba 2010, Tunisia
e-mail: fatma.hmissi@ensi-uma.tn

Sofiane Ouni
*Cristal Laboratory, ENSI*
*Manouba University Campus*
Manouba 2010, Tunisia
e-mail: sofiane.ouni@insat.rnu.tn

*Abstract*—Nowadays, all sectors utilize devices that are part of the Internet of Things (IoT) for the purpose of connecting and exchanging information with other devices and systems over the Internet. This increases the diversity of devices and their working environments, which, in turn, creates new challenges, such as real-time interaction, security, interoperability, performance, and robustness of IoT systems. To address these, many applications protocols were adopted and developed for devices with constrained resources. This paper surveys communication protocols divided according to their goals along with their merits, demerits, and suitability towards IoT applications. We summarize the challenges of communication protocols as well as some relevant solutions.

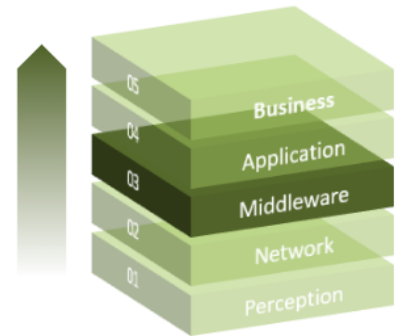

Figure 1: 5-Layer Architecture for IoT.

## I. INTRODUCTION

The Internet of Things (IoT) [1], [2] refers to physical things that have been combined with sensors, actuators, and technologies in order to exchange data with other devices and systems on the network. The IoT is used to make people's lives and businesses easier in many areas. Generally speaking, there is no standard architecture for the Internet of Things systems, but what is certain is that all architectures are composed of several parts which interact and communicate with each other without human intervention. The Internet of Things connects the real world of things to the virtual world of networks and the Cloud.

Currently, there is no universally reference architecture for the IoT. There is a considerable amount of architectures available for IoT systems. The most widely architectures [3]–[6] used for IoT solutions are: 3-layer, Service-oriented Architecture (SoA) and 5-layer architectures. However, the 5-layer architecture, also called middle-ware architecture, presents the very common architecture in IoT. As shown in Figure 1, middle-ware architecture divides IoT systems into 5 layers: Perception, Network, Middle-ware, Application and Business. The Perception layer, or the device layer, represents the physical level objects, having as main function the gathering of useful information from the surroundings. Here, a number of sensors and actuators are used to monitor - control the physical objects. The Perception layer transmits then the gathered information to Middle-ware layer using the Network layer. The Network layer, or communication layer, connects Perception layer and Middle-ware layer by transporting data provided by Perception level to Middle-ware layer. The Network layer in IoT architecture does not present Network layer of the ISO/OSI model, that routes data within the network only along the best way. The Network layer of IoT 5 architecture includes all the technologies and protocols that make the connection possible between the Perception and Middle-ware layers. The Middle-ware provides some advanced functionalities, such as storage, processing, aggregation and filtering of data. The Application presents the collected and analyzed information to the end user. Lastly, the Business layer enables systems' administrators to manage and supervise the IoT platform's entire functionality.

Figure 2 introduces a typical scenario of the IoT system where the interaction between the different parts is clearly presented. The IoT devices [7] are physical things. They are equipped with embedded sensors, actuators, and controllers to interact with the physical environments to collect information or to change the actual status. A device can exchange data either with other devices or with data-center, the Cloud, or other servers. The Gateway represents a physical entity that is composed of several electronic devices. The main purpose of the Gateway is to connect to different networks having different typologies. It contains software that translates the protocols to establish communication between the things and the network. The number of connected devices is expected to grow rapidly, with a predicted 75 billion devices worldwide expected to be connected to the Internet by 2025 [8]. This great number of connected devices is expected to generate unlimited data. As a result, an enormous amount of data to be stored, processed, and made available in a continuous, efficient, and easily interpretable manner is growing rapidly, which puts a lot of pressure on the Internet infrastructure. To solve this problem, companies combined the capabilities of IoT and cloud computing. The technology of cloud computing assists in alleviating the pressure on the Internet infrastructure by storing, processing, and transferring data to the Cloud instead of to the connected devices. Many platforms, called IoT Cloud Platforms, exploit Cloud Computing features to provide IoT services. For this purpose, a number of open sources and proprietary IoT platforms have been proposed and implemented to provide many efficient and easy IoT services, such as data collection, storage, analysis, monitoring, control, and management of connected things. Today, more than 300 IoT platforms are available on the market [9]. Mobile and Web applications make the IoT very user-friendly. A mobile application is a software application that is created to run on mobile devices especially those that are small and wireless. A Web application is a software application that is hosted on a server and accessible through a Web browser. Mobile and Web applications allow users to perform a set of specific functions and tasks on the Internet. These functions and tasks are summarized in the connection, monitoring, control, and management of connected objects.

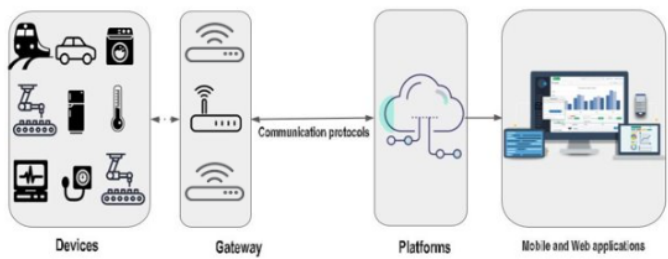

Figure 2: IoT Basic Scenario.

Communication Protocols are used to allow the connectivity for data exchange between physical or virtual entities, e.g., Devices and the Cloud, by defining rules and constraints where several requirements must be taken into account by these protocols when building IoT systems.

In this paper, we focus on communication protocols designed for IoT. Communication protocols are subdivided into three types, namely, (i) messaging (ii) device management, and (iii) service discovery.

The weaknesses of the current solutions have motivated the improvement of the existing protocols that seek to enhance the Internet of Things ecosystems' performance and avoid faults.

Thus, this work aims at presenting an extensive survey about the existent communications protocols that can be used in IoT applications. Different from the current existing surveys in the literature, this work does not only consider existing and well-known base protocols, but also all relevant solutions that have been introduced during recent years. The main contributions of this paper are summarized as follows:

- Classifies communication protocols into three groups, namely messaging, device management, and service discovery. Sequentially, we define each group.
- Surveys the most common messaging protocols used in IoT solution.
- Outlines the usage level of messaging protocols
- Overviews of device management and service discovery protocols.
- Identifies the problems most studied by the existing protocols in IoT scenarios.
- Reviews of the studied solutions that improve existing protocols.

The remainder of this paper is organized as follows: Section II introduces a taxonomy of communication protocols. Section III presents the messaging protocols used in IoT applications. Section IV outlines the usage level of the IoT messaging protocols and conclude the most used protocol. Section V lists the communication protocols for device management. Section VI sums up the communication protocols for service discovery. Section VII introduces the challenges for the communication protocols and surveys the recent approaches to the protocols enhancement. Section VIII concludes the paper.

II. TAXONOMY OF IOT APPLICATION PROTOCOLS

As introduced in section I, the Network layer of IoT architecture enables IoT devices to communicate with Middle-ware layer, by including several protocols. A protocol represents the rules and formats that IoT devices use to establish connections with Middle-ware layer. The Network layer's protocols are built on a stack of protocols [10]. Figure 3 shows a list of some of the most commonly used protocols, organized according to the TCP/IP paradigm.

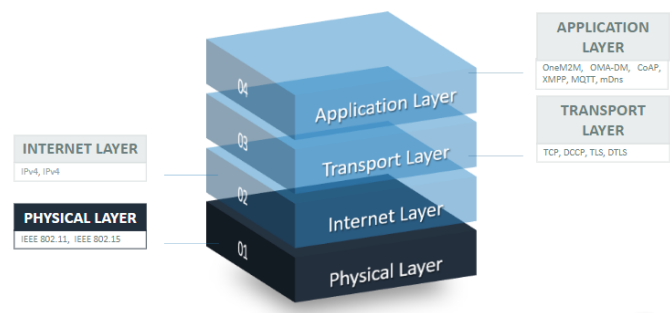

Figure 3: IoT Protocol Stack.

However, in this paper we focus on Application layer protocols of TCP/IP paradigm. The Application layer provides protocols that enable device to send and receive information, and should not be confused with the Application layer of the IoT 5-layer architecture. A few examples of Application layer protocols are: Message Queuing Telemetry Transport (MQTT) [11], [12], Constrained Application Protocol (CoAP) [13], [14], Data Distribution Service (DDS) [15]–[18].

As shown in Figure 4, those protocols can be classified into three different groups. Specifically, application protocols can be classified by their purpose: messaging protocols, device management protocols, or service discovery protocols.

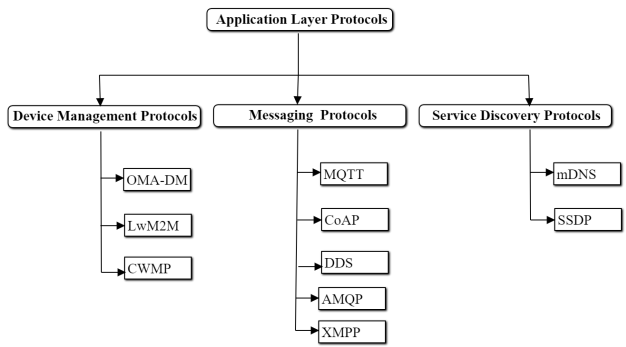

Figure 4: Taxonomy of Application Protocols for IoT.

*a) Messaging Protocols:* It define the formats, rules and functions for transferring messages between the components of a IoT system system. It is possible to build an IoT system with the typical messaging protocols based on classical HTTP Web requests even if they do not have certain requirements. However, they are no longer the right choice for Internet of Things. A IoT messaging should has important characteristics, namely speed (the amount of data that can be transmitted per second), latency (the amount of time needed to transmit a message), power consumption and security. New messaging protocols have been implemented, such as MQTT, CoAP, XMPP.

*b) Device Management Protocols:* A huge number of connected devices are deployed in remote, hostile and hard to reach places, which makes their configurations and maintenance difficult. Many solutions are proposed to provide device management necessity. For example, Perumal et al. [19] proposed a lightweight IoT device management framework for smart home services. Also, various protocols [20], called device management protocols, are proposed to ensure IoT network management. Here, a device management protocol [21] provides device location and status information to adapt the topology of IoT networks. A device management has advanced functionalities, such as disconnect and locate lost devices, modify security settings, delete device data.

*c) Service Discovery Protocols:* Mechanisms for discovery are important to use the services of the Internet of Things. Service discovery is a process of automatically locating the appropriate services. Ahmed et al. [22] proposed a secured service discovery technique for the Internet of Things. Several protocols [22] are proposed to handle service discovery. A service discovery protocol locates services across widely distributed and heterogeneous networks that are relevant to an entity of interest in the real world.

## III. MESSAGING PROTOCOLS

IoT cannot rely on a single protocol for all needs [23]. Consequently, several of available messaging protocols are chosen for various types of requirements of the IoT system [24]. Thus, in the rest of this section, the most relevant protocols are cited with their descriptions.

*a) Message Queuing Telemetry Transport (MQTT):* It is a lightweight [11] [12] and flexible [25] messaging protocol. MQTT uses different approaches for routing mechanisms, such as one-to-one, one-to-many, or many-to-many, making the connection between IoT and Machine-to-Machine (M2M) to connected devices/applications possible [25]. M2M is used to provide communications between machines without human intervention. MQTT is designed as a publish-subscribe model [25], using TCP as transport layer protocol.

Figure 5 shows the process of message exchange in MQTT. MQTT consists of multiple clients connected to a central broker, which is a server running somewhere in the Internet network [26]. These clients can be a publishers or subscribers. A publisher is producer that publish messages on a particular topic [25]. However, a subscriber is a consumer that subscribes to receive published messages on a topic. Every time the MQTT Broker gets a new publish message to a specific topic, it broadcasts this message to the interested subscribers.

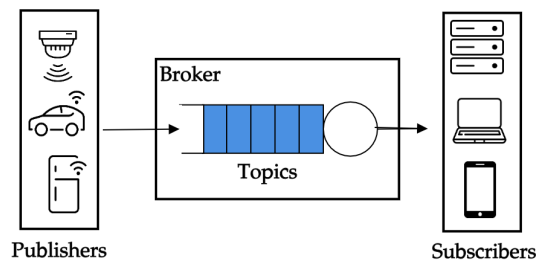

Figure 5: MQTT Architecture.

The MQTT protocol has several advantages [27], [28]. First, the messages may be sent/received at anytime, hence the communications are asynchronous. Second, the method of communication used is very simple. Third, MQTT provides the reliability of messages by providing 3 level of QoS [29] . Here, the publisher uses a QoS level for each published message to ensure that the data reaches its recipient. In the following, we present the QoS level supported by MQTT:

- QoS 0 (At most once): The receiver does not confirm the reception of messages, and the producer does not wait for such confirmations.
- QoS 1 (At least once): if the publisher uses the level 1 than it guarantees that the data is delivered at least one time to the receiver. For this purpose, the receiver confirms the delivery of data to the publisher, the publisher store the sent data to re-transmit it if necessary.

- QoS 2 (Exactly once): if the publisher uses the level 2 than it guarantees that the data is delivered at exactly one time to the receiver.

However, MQTT has some limitations. We will address some of them. The MQTT protocol is used between devices and Cloud, but it is not commonly used between devices. Another disadvantage is that MQTT uses TCP/IP and the use of TCP/IP requires more communication. The last drawback concerns the usage of a broker. A broker has restricted communication capabilities, and all nodes are connected to it. As a result, when the broker fails, the communication breaks down.

*b) Constrained Application Protocol (CoAP):* CoAP [13] [14] is mainly used in a constrained environment with constrained devices and constrained networks. As Figure 6 shows, CoAP environments use unicast and multi-cast request-response model for interaction between multiple clients and multiple servers by sending request and response messages using a URI with GET, POST, PUT and DELETE actions over UDP to keep things lightweight.

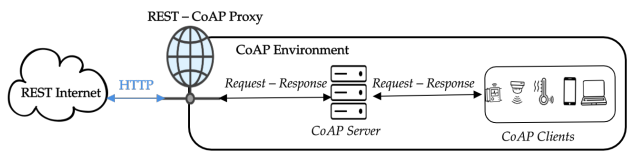

Figure 6: CoAP Architecture.

CoAP uses two modes of messages for request-response communication, namely piggyback and separate. The main difference between these two modes resides in the time and the way of responding [27], [28]. In direct communication between the client and server, piggyback means that the server sends its response message directly after receiving the request. In this case, the server response comes with an acknowledgment (ACK) message. While separate mode is used in indirect communication between client and server, the response is sent in a separate message from the ACK and may take some time for the server to provide it.

Also, CoAP provides two types of messages for the reliability and duplication of messages. The two types of messages are confirmable and non-confirmable. These two messages are used, respectively, for reliable and unreliable transmission. The use of the confirmable message requires the use of an ACK message to confirm the message's arrival, while the use of the non-confirmable message requires no use of an ACK message. The main merits of the CoAP protocol are that it can be used with constrained devices in interaction device-to-device, and that it allows fast communication since small packets are sent with the UDP layer. This protocol cannot be used in asynchronous communication because it does not support publisher-subscriber architecture. Also, it does not support broadcast. The clients cannot use a topic to send and respond to messages.

*c) Data Distribution Service (DDS):* [15]–[18] is used for real-time and industrial M2M communications, running over both TCP or UDP. DDS supports broker-less architecture where it uses a publish-subscribe model for interaction between entities without the use of a Broker. The tasks of a broker are handled by Data Writers (DW) and Data Readers (DR). The main advantages of DDS protocol are that the data usage is fundamentally anonymous, since the publishers do not enquire about who consumes their data, and the probability of system failure is limited (system more reliable) because there is no single point (no broker) of failure for the entire system [15]. The most remarkable disadvantage of DDS is that it is designed for Industrial application (IIoT) with considerable hardware resources. This makes the implementation for constrained devices that need a Lightweight protocol even harder. The other disadvantage is related to the increase of the communication workload by the publishing of data even if there are no interested subscribers [15].

*d) Advanced Message Queuing Protocol (AMQP):* AMQP [15], [16] is designed as a publish-subscribe model, which uses TCP as transport layer protocol. Mainly as described in Figure 7, it has three components, Publishers, Subscribers and, both parts of an AMQP Broker are Exchanges of Message queues. The Publisher creates a bare message and sends it to the Exchanges components that are used to forward the messages to appropriate message queues using the routing keys contained in messages. The latter can be stored into message queues before forwarding them to Subscribers. If there are more subscribers interested in a particular message, the broker can duplicate the messages and send their copies to multiple queues waiting for annotated messages from subscribers. The main advantage of the AMQP protocol is that it could be used in device-to-device, device-to-Cloud, and Cloud-to-Cloud interaction. But its main disadvantage is that the publishers and subscribers cannot publish and subscribe using the topic.

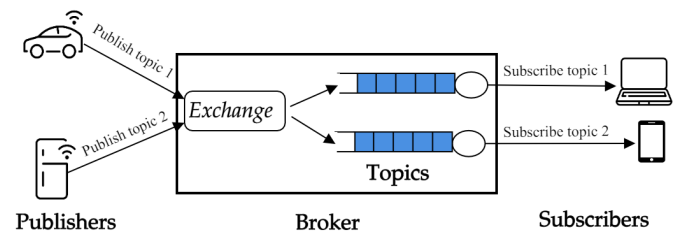

Figure 7: AMQP Architecture.

*e) eXtensible Messaging and Presence Protocol (XMPP):* As introduced in [30]–[32], XMPP, also known as Jabber, is a standard initially designed for instant messaging and exchange of messages between applications no matter which operating system they are using in IoT. It is designed to allow users to send messages in real-time and manage the presence of the user. XMPP supports Publish-Subscribe and Request-Response models with TCP transport protocol. To exchange messages between clients and servers XMPP uses streams of stanzas. XMPP is a text-based protocol where XMPP stanzas [30] [31] [32] are Extensible Markup Language (XML) messages exchanged between clients. The main advantage of the AMQP protocol is that could be used in device-to-device, device-to-Cloud interaction. As AMQP, DDS, and CoAP, topics are not used to publish and subscribe with XMPP.

The discussed IoT messaging protocols have similarities and differences among a number of features [33], [34].

However, all the cited protocols lack of IoT device management and service discovery procedures. When building an IoT system, it is important to think about the protocol's characteristics to ensure that it meets functional and operational needs. For that purpose, we provides a comparative analysis (see Table I), where the differences and similarities in the relationships between messaging protocols are clear. In the comparative analysis, we considered key features such as pattern, QoS, Payload's format and maximum size.

## IV. MESSAGING PROTOCOLS TRENDS

This section investigates the level of use of IoT messaging protocols. First, we will analyse the results of the annual survey realized by the Eclipse Foundation [35] between 2016 and 2018. Then, we will be interested in the support of the messaging protocol by IoT Cloud Platforms.

Figure 8 adopts the evaluation of the use of the messaging protocols in IoT systems. According to the results of this survey [35], MQTT and HTTP are the two most used protocols. This survey confirms that MQTT is the choice for IoT solutions since it is the denominator by 62.61%. While HTTP usage is declining to 54.10 %, this could be due to the advantages of using the light-weight version of HTTP (HTTP/2). In addition, the AMQP protocol has significant traction in terms of its usage in 2018 compared to 2017 (18.24%). Furthermore, the use of the new WEBSOCKETS protocol shows a very high usage level of 34.95%.

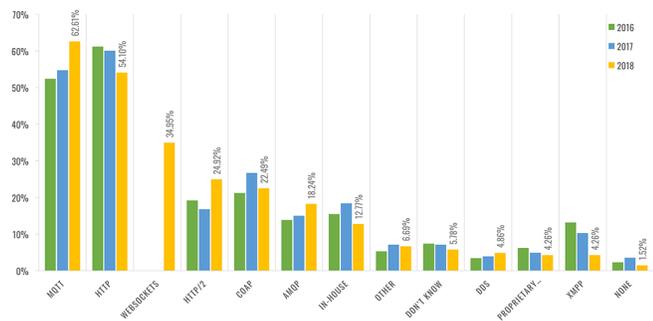

Figure 8: IoT Developer Survey Results Showing the Trend in the Usage of Messaging Protocols Between the Years 2016 and 2018 [35]

We will now examine the level of use of messaging protocols by examining how messaging protocols are supported on the IoT Cloud platforms.

Today, IoT networks transmit more data than they can handle effectively because of the amount of data generated and exchanged by devices. This behavior negatively affects the performance of IoT applications, such as increased response time and loss of network connectivity. On that basis, several cloud platforms were offered to improve IoT applications. Simply put, these platforms are designed to reduce network congestion. IoT communication protocols (messaging protocols) play an important role in IoT applications. Messaging protocols provide the ability to transfer data between devices and cloud platforms. However, not every messaging protocol can be used between devices and cloud platforms. Each IoT cloud platform supports its own specific messaging protocols. Here, and throughout the remainder of this section, we look at the support levels of the messaging protocols across the cloud platforms. But above all, we're going to outline three of the major existing IoT Cloud platforms.

Microsoft Azure IoT [36]–[38] Suite is a cloud computing PaaS that allows developers to publish web applications running on different frameworks and written in different programming languages, such as any. NET language, node. js, PHP, Python, and Java [1]. The IoT devices send the data to the cloud gateway directly or via a gateway, depending on the network capabilities of the IoT device. The Azure IoT Suite is used to build Internet of Things systems and applications by gathering, storing, processing, managing and analysing data. The processed data can later be delivered to other business and presentation applications.

IBM BlueMix [36], [39] is a Cloud-based PaaS developed by IBM for building, running, deploying and managing applications of all types such as web, mobile, new smart devices, and much more, runs on Soft-layer that is IBM's worldwide IaaS. BlueMix used to develop apps in many programming languages, including 1) JavaScript to develop mobile apps in iOS, Android, and HTML, 2) Node. js, Ruby, PHP, Java, Go, and Python and many more to develop web apps.

AWS IoT [36]–[38] is a Cloud-based PaaS developed by Amazon for facilitating security, services, and support. AWS IoT used to create apps in many languages, such as PHP, .Net, Node. js, Ruby, PHP, Java, Go, and Python and many more, to develop web apps or Docker containers that run on an application server with a database. The main services of this platform are: device management, rules and analytics, and data storage and integration, as well as security.

Table II presents the lists of the messaging protocols supported by the IoT Cloud platform for data transmission. Table II confirms that the MQTT protocol takes first place in the IoT market since it is supported by all the IoT cloud platforms. HTTP comes in second, followed by WebSockets.

Based on Figure 8 and Table II, our in-depth examination of messaging protocols leads us to the conclusion that MQTT is the messaging protocol with the greatest impact on the IoT. The MQTT protocol is the lightest, the most reliable, and the one that has the least overhead.

Table I: Features of Messaging Protocols.

| Feature | Messaging Protocols | | | | |
|---|---|---|---|---|---|
| | MQTT | DDS | CoAP | AMQP | XMPP |
| Pattern | Publish-Subscribe | Publish-Subscribe | Client-Server Publish-Subscribe | Publish-Subscribe Publish-Subscribe | Publish-Subscribe Publish-Subscribe |
| Transport | TCP | TCP, [UDP] | UDP, [TCP] | TCP, [UDP] | TCP |
| Scope | Device-to-Cloud Cloud-to-Cloud | Device-to-device Device-to-Cloud Cloud-to-Cloud | Device-to-device | Device-to-device Device-to-Cloud Cloud-to-Cloud | Device-to-cloud Device-to-Cloud Cloud-to-Cloud |
| QoS Level | 3 | 23 | 2 | 3 | none |
| Addressing | Topic | Topic Key | URI | Queue Queue Routing key | Jabber Identification |
| Max Payload Size | 256 MB | 64 KB, 4 GB with block-wise transfer | 40 B–1KB (without IP fragmentation), 1 MB–1GB with block-wise transfer | defined by end-points | defined by end-points (64 KB stanza size) |
| Payload Format | Arbitrary | Strongly defined types, Mixed | JSON,XML | - | XML |

## V. DEVICE MANAGEMENT PROTOCOLS

A huge amount of heterogeneous devices, which are integrated into IoT, need to be (re)discovered, reconfigured, and maintained to fix security issues, deploy new features, or recover from their failures. It is possible to manage devices with the IoT messaging protocols by inventing new building blocks. It seems that these protocols are no longer the right choice for device management because of the high cost of development, where for every new management feature, a new block should be developed. To solve this problem, new protocols known as Device Management Protocols are proposed. A device management protocol enables the abstraction of an IoT/M2M device as a managed object to make the management of the device much easier [50].

Open Mobile Alliance Device Management standard [51] [52] [53] named as OMA-DM and designed by Open Mobile Alliance for device management, is used for Terminal M2M devices and Mobile terminal devices, e.g, Mobile phones, Smartphones, Tablets, laptops. Mobile network operators and enterprises use OMA-DM to manage mobile devices remotely. The main features of OMA-DM are: read and write configuration or monitoring nodes, read and set parameter keys and values, Firmware Update Management Object (FUMO), software components management object (SCMO) that means install, upgrade, or uninstall software elements. OMA-DM has several demerits. The OMA-DM protocol is designed only for no constrained and fixed devices. Another disadvantage is that OMA-DM cannot be used for industrial applications and cannot be built on top of the MQTT protocol. The disadvantage before the last is that it supports only XML serialization format and it does not support either Binary, Plain text, or TLV and JSON serialization format. The final disadvantage refers to the no support of interoperability.

Lightweight M2M [54] [55] (LwM2M) is a client-server standard developed by the Open Mobile Alliance (OMA). It is an OMA-DM successor. The LwM2M is a standard device and service management built on top of CoAP to ensure remote management and configuration of constrained and powerful devices. It can beneficiate from efficient communication in M2M and IoT environments over UDP and SMS bearers. So, SMS can be used for waking up the device or any GET, POST, and PUT request. The LwM2M main features are: device monitoring and configuration, server provisioning (bootstrapping) and firmware upgrades. There are numerous advantages of the LwM2M protocol. The most remarkable advantage of LwM2M is that it could be used with fixed and mobile-constrained devices. Another advantage refers to the support of the industrial application and interoperability. The most important disadvantages of LwM2M are: cannot support XML serialization format, cannot be built on top of MQTT and cannot be used in telecommunication applications.

The Broadband Forum defined CPE WAN management protocol (CWMP) that is used for remote management of home and business network devices, such as modems, gateways, routers, and VOIP phones (see Technical report 069 [56] [57] known as TR-069). The main capabilities of this protocol are firmware management, auto-configuration, dynamic service provisioning, software module management, status monitoring, performance monitoring, and diagnostics. The TR-069 uses SOAP (Simple Object Access Protocol)/HTTP protocol for communication between network devices called the Customer Premises Equipment (CPE) and central server called the Auto-Configuration Servers (ACS). The CPE and ACS present the main components of this protocol. TR-069 has the same disadvantages as OMA-DM.

IoT devices management protocols are not oriented for communication and service discovery features. Our depth study allows us to conclude that the device management protocol with the greatest impact on the IoT is the LwM2M. The overhead, footprint, and server load of the LwM2M are lighter than TR-069 and OMA-DM protocols, while the response time of LwM2M is faster than TR-069 and OMA-DM protocols.

Table II: Messaging protocols supported by existing IoT platform for data transmission.

| IoT Platform | Messaging Protocols | | | | | | |
|---|---|---|---|---|---|---|---|
| | MQTT | HTTP | WebSockets | HTTPS | CoAP | AMQP | XMPP |
| Microsoft Azure IoT [36]–[38] | ✓ | ✓ | | ✓ | | ✓ | |
| IBM BlueMix [36], [39] | ✓ | ✓ | ✓ | ✓ | | | |
| AWS IoT [36]–[38] | ✓ | ✓ | | ✓ | | | |
| Kaa [37], [40] | ✓ | ✓ | | | | | |
| DeviceHive [41] | ✓ | | ✓ | | | | |
| ThingSpeak [42]–[44] | ✓ | ✓ | ✓ | | | | |
| OpenMTC [45] | ✓ | ✓ | | | | | |
| SiteWhere | ✓ | | ✓ | | ✓ | ✓ | |
| Linksmart [40], [45], [46] | ✓ | | ✓ | ✓ | | | |
| OpenRemote [45] | ✓ | ✓ | ✓ | | | | |
| Zetta [39] | | | ✓ | | | | |
| Google IoT Core [47] | ✓ | ✓ | | ✓ | | | |
| Oracle IoT [48] | ✓ | ✓ | ✓ | ✓ | ✓ | ✓ | ✓ |
| Cisco Kinetic [49] | ✓ | ✓ | ✓ | ✓ | | ✓ | |
| Eclipse Homo | ✓ | ✓ | | ✓ | ✓ | ✓ | |

## VI. SERVICES DISCOVERY PROTOCOLS

Service Discovery Protocols (SDPs) are communication protocols that provide mechanisms to help clients to discover services available on the network. There are several SDPs for the IoT environment. This section focuses on the most known SDPs by introducing the following protocols: mDNS, SSDP. Multicast Domain Name System (mDNS) [58] [59] [60] is an open protocol defined by IETF, which requires minimal configuration, based on the Internet Protocol (IP) and the User Datagram Protocol (UDP). An mDNS client can discover a thing's endpoint by resolving its hostname to an IP address. An mDNS client has to send an IP multi-cast query message over the network. The message calls the host with that name to reply and identify. Once the host receives the message, it replies via a multi-cast message that contains its IP address. All nodes in the network receiving that multi-cast message update their mDNS caches accordingly. This protocol, coupled with DNS-based Service Discovery (DNS-SD), offers the flexibility required by environments where it is necessary to automatically integrate new devices and perform DNS-like operations without the presence of a conventional DNS server.

The Simple Service Discovery Protocol (SSDP) [58] [59] [60] is an open protocol, based on IP, UDP, and SOAP [58] [59] [60]. An SSDP client discovers SSDP services by multi-casting a discovery request to the SSDP multicast channel and port. SSDP services listen on that channel until they receive a discovery request that matches the service they offer, then they respond using a unicast response. This protocol—included in the Universal Plug-and-Play (UPnP) architecture—makes it possible to transparently plug and play devices without the need for any manual configuration.

## VII. CHALLENGES AND ENHANCEMENTS OF COMMUNICATION PROTOCOLS

IoT protocols have limitations and drawbacks. Among these, we highlight communication protocols challenges:
- Real-time and industrial communication issues.
- Not suitable for constrained devices.
- Interoperability issues.
- Security issues.
- Quality of Service (QoS) issues.

Motivated by the presented issues, several new solutions have emerged recently. In this section, an overview of studies focusing on the improvement of existing and well-known base protocols are divided and presented according to their proposals. Table III summarizes the existing studies of some widely efficient and recently enhanced approaches for application layer protocols in IoT environment.

*a) Real-Time and Industrial Communication:* Several applications in IoT fields, such as medical, factory, and transportation are time-sensitive applications. Mostly, the delays of communications between the different parts of the IoT systems are in-bounded. Therefore, the real-time requirement is one of the challenges of communication protocols. Most IoT solutions involve time constraints to gather and process information, make decisions, and deliver actions that system components must perform. When time restrictions are present, the system is said to be real-time if at least one of the tasks is performed but it must be executed before a certain deadline.

XMPP and DDS protocols are designed for real-time communication. Even though the other protocols, such as MQTT and CoAP, have received a lot of attention due to their simplicity and scalability, none of them support real-time interactions.

To address this, many approaches are proposed to add enhancement to applications protocols without changing their simplicity and scalability. Kim et al. [61] propose to integrate MMS and MQTT protocol for Internet of Things industrial applications. Konieczek et al. [63] presented a lightweight Java implementation of the Constrained Application Protocol called jCoAP that enables CoAP-based communication for embedded devices with comparably small latencies (real-time interaction).

*b) Constrained Devices:* IoT devices are constrained. They have limited capabilities, memory, and energy. And the use of heavy communication protocols on these devices reduces the performance of IoT communication. i.e., shut down the devices quickly, increase the delay of communication. Power consumption is one of the most constrained aspects of IoT devices, which makes the most powerful applications protocols not suitable for Internet of Things ecosystems.

Table III: Surveys on Communication Protocols Challenges and Enhancement.

| Challenge | Focus | Protocol | References |
|---|---|---|---|
| Real-time communication | Industrial application | MQTT | [61] |
| | IoT based system | MQTT | [62] |
| | Embedded devices | CoAP | [63] |
| | Prototype Medical Instruments Applied to Neurodegenerative Disease Diagnosis | MQTT/AMQP | [64] |
| Constrained devices | Power saving | MQTT | [65], [66] |
| | Power saving | MQTT-SN | [67], [68] |
| | Power saving | CoAP | [69] |
| | Decrease the computational complexity of the clients | MQTT | [70] |
| Interoperability | Technical Interoperability | MQTT/HTTP | [71] |
| | Technical Interoperability | MQTT | [72] |
| | Syntactical interoperability | All protocols | [73] |
| | Semantic interoperability | MQTT | [74] |
| Security | Authentication | MQTT/MQTT-SN | [75]–[79] |
| | User authority to information access | MQTT | [80] |
| | User Registration | MQTT | [81] |
| | Denial-of-sleep attacks | CoAP | [82] |
| Quality of services | Control the traffic flow between the subscribers and publishers | MQTT | [83], [84] |
| | Maintain message order | MQTT | [85] |
| | Transit urgent message first | MQTT | [86] |
| | Reduce the delivery of unnecessary messages | MQTT | [87] |
| | Data Delivery in Mobile Scenarios | MQTT | [88] |
| | Network Congestion Control | CoAP | [89] |
| | Object Discovery | CoAP | [89] |

Although MQTT is a lightweight protocol, it has its drawbacks for extreme environments. MQTT clients must support TCP and would normally keep an open connection to the broker at all times where packets loss and connection drop rates are high or computing resources are scarce. Moreover, topic names are often long ones, which make the header bigger and use significant bandwidth and power as well. To address this, many variations and enhancements are proposed. First, Query Telemetry Transport for Sensor Networks (MQTT-SN) was created [65], which runs over UDP. UDP is mainly used for sensor nodes and devices with low computing performance. MQTT-SN requires additional gateways to connect the clients to the MQTT broker over UDP, which can be suitable for devices with multicast support. A modification to MQTT-SN with additional security elements adopted from DTLS is proposed to replace the DTLS protocol to enable shorter lightweight packet headers [15].

Akintade et al. [66] proposed another architecture to facilitate the development of energy-efficient and low-cost IoT solutions, namely, the aMQTT architecture. The architecture is based on the existing MQTT architecture and the low cost ESP8266 IoT hardware platform. Second, many enhancement solutions were added to the MQTT-SN protocol to increase its performance especially in extremely lossy channels where re-transmission creates a huge overhead in terms of power consumption, delay, and processing. Alshantout et al. [67] created MQTT-SN with LT (MQTT-SN-LT). They aim to use Luby Transform Codes (LT) with the MQTT-SN-QoS1 protocol without changing the protocol itself. The authors [68] proposed to add Network Coding to an MQTT-SN network.

*c) Security:* Application protocols were not designed with security in mind [90]. They are based on common security solutions, such as DTLS and TLS which are not sufficient for optimal security as they reduce the performance of IoT systems. To go further, these solutions are very heavy for constrained devices. Added to that, certain attacks are no longer covered by these solutions which require the development of new standards to improve the security levels of each protocol. In the rest of this section, several attacks and problems are cited as well as their solutions. The flow of the distributed messages between the users of application protocol based on Publisher-Subscriber models is insecure. Wherein authentication layer authenticating credentials are sent in plain text and some form of encryption should be used. In authorization layer all users connected to the broker are listening to a Topic and receiving all the information.

For authentication layer several works are proposed such as the works [75], [77]–[79]. Blockchain is a distributed immutable time-stamped ledger. Today, researchers are combining the blockchain and the IoT together to increase the security level of IoT applications. In this context, new security schemes based on blockchain technology are pro-posed, for example [77]–[79], where M. Abubaker et al. [77] proposed a lightweight authentication schema for the MQTT protocol based on blockchain, and also F. Buccafurri et al. [78], [79] proposed a lightweight OTP(One-Time Password)-authentication schema based on blockchain for the MQTT protocol. ChaCha20- Poly1305 AEAD solution is proposed as a lightweight security scheme for MQTT/MQTT-SN communication in [75]. Since, in a MQTT environment, a user in the broker's access is authorized to access all information, after their connection to the broker the user is listening to a Topic and receiving all the information. A new solution of certified authority is opted for in [80] to generate two kinds of certificates, the first one for the client and the second one for the Topics. One of the most well-known types of attacks in network sensors is the Denial of Sleep (DoS) attack [91], [92]. DoS prevents the radio from reaching sleep mode, which would entirely consume the battery. In normal working conditions, sensors' energy consumption ratio consumes their batteries over months, however a denial of sleep attack drains them over days by keeping the radio transmitter system on the sensor nodes active. So, DoS attacks aim at depriving victims of devices entering low-power sleep mode. Since the CoAP protocol suffers from this type of attack where Internet-located attackers can force IoT devices that run CoAP servers to expend much energy by sending lots of CoAP messages to them, a new solution is proposed by adding a block to filter the CoAP messages en route before entering the network [82].

*d) Interoperability:* Interoperability is meant to make communication among heterogeneous devices and software applications from different vendors possible. Interoperability has four dimensions: technical, syntactical, semantic, and organizational interoperability [93]. There is no compatibility in inter-communication between application protocols. Messages are not supposed to be exchanged. Thus, we need new standards to convert communication protocols and to enlarge the protocol's capabilities for larger interoperability.

A new efficient application layer gateway that converts MQTT messages into HTTP is proposed in [71]. To address the problems of the interconnection of embedded systems in networks, the authors of [72] aim to dynamically model and create links between MQTT brokers based on multi-agent systems to establish the highest level of connectivity for brokers to ensure maximum transmission of messages to subscribing clients. Since there is no compatibility between the sensors, where each sensor for example has its own data display units, there is a need for common semantics for these sensors. To solve this problem, several standards have been developed to ensure that the precise meaning of exchanged information can be understood by any other application that was not initially developed for that purpose. A semantic data extraction implementation over MQTT for Internet of Things centric wireless sensor networks was introduced [74].

*e) Quality of Service (QoS):* The QoS characterizes the quality of communication links between nodes. Generally, it is the capacity to carry the traffic between nodes in the best condition, such as in terms of availability, packet loss rate, and throughput. So, to ensure good communication it is recommended to define clearly the quality metrics and to enhance the communication protocols accordingly. Quality of Service is the strength of application layer protocol, that represents the ability to configure the performance and reliability of the network. Some protocols do not define any QoS level which reduces their performance, while others, such as CoAP, MQTT, and DDS define different levels of QoS which address different requirements, such as message delivery, timing, loose coupling, and fault tolerance. As MQTT provides only three levels of QoS for different classes of traffic, so many drawbacks arise.

Firstly, the traffic flow between subscribers and publishers is not controlled since publishers send data to broker and broker forwards it to subscribers which could increase the number of packet losses and delays. A new flow control mechanism is designed to overcome the flow control problem of MQTT where the publisher can overwhelm the subscriber [83], [84].

Secondly, MQTT does not support the urgency of the message. Hence, normal and urgent messages are processed with the same priority. Many approaches are designed for this purpose. Hwang et al. [86] proposed a new method to expand the functions of the MQTT to transmit urgent messages first by creating a U-Mosquitto broker capable of processing urgent messages.

However, MQTT protocol has vulnerability to maintain order between messages, which is very important in some home automation, such as controlling gas valve. Hwang et al. [85] designed and implemented a reliable message transmission system using MQTT protocol to maintain messages order.

Also, the absence of a standard for controlling the number of messages received is such a serious problem where the subscriber devices are forced to receive all messages even if they do not need to receive them frequently. To solve this problem, reducing the delivery of unnecessary messages is the best solution. Hwang et al. [87] focused on the MQTT protocol that is currently used to deliver messages between IoT devices and proposed the concept of Reception Frequency Control (RFC), which is designed to control the frequency at which subscribers receive messages.

## VIII. CONCLUSIONS

Application communication protocols in IoT ecosystems are used to successfully interact between IoT devices and servers / Clouds that process the information collected. Application protocols specific to IoT have been developed to meet the requirements of devices with limited resources, and those of networks with low bandwidth and high latency. However, establishing low-cost communications is not enough. These protocols must allow data to be exchanged and this data must be understood by the entities of different types which receive them. The interoperability of distributed applications is defined as the ability of success for the IoT thanks to a set of application protocols for users to communicate and exchange data and services, wherever they are in the world regardless of the origin of the equipment they use.

There several challenges in front of IoT application protocols. These challenges are related to the drawbacks of application protocols. Those challenges can be summarized in the following points: not suitable for real-time and industrial application, not suitable for constrained devices and lack of interoperability, security mechanisms and Quality of Service (QoS).

In this paper, we surveyed the most suitable communication protocols for the Internet of Things and related challenges of IoT issues by introducing relevant and recent approaches for improving the performance of application layer IoT systems.

The studied application protocols, in general, are based on MQTT, CoAP applications protocols. This is justified due to MQTT and CoAP being already the most suitable solutions in IoT since they are better suited to the application layer criteria: message size, overhead, power consumption, resource requirement, bandwidth, and reliability.

REFERENCES

[1] H. Fatma and O. Sofiane, "A review of application protocol enhancements for iot," The Fifteenth International Conference on Mobile Ubiquitous Computing, Systems, Services and Technologies (UBICOMM 2021), pp. 7–13, 10 2021.
[2] A. Syed, D. Sierra-Sosa, A. Kumar, and A. Elmaghraby, "Iot in smart cities: A survey of technologies, practices and challenges," Smart Cities, vol. 4, pp. 429–475, 03 2021.
[3] M. Lombardi, F. Pascale, and D. Santaniello, "Internet of things: A general overview between architectures, protocols and applications," Information, vol. 12, p. 87, 02 2021.
[4] L. Antão, R. Pinto, J. P. Reis, and G. Gonçalves, "Requirements for testing and validating the industrial internet of things," 04 2018.
[5] O. Ali, M. K. Ishak, M. K. Bhatti, I. Khan, and k. kim, "A comprehensive review of internet of things: Technology stack, middlewares, and fog/edge computing interface," Sensors, vol. 22, p. 995, 01 2022.
[6] N. Abdulkareem, S. Zeebaree, M. M.Sadeeq, D. Ahmed, A. Sami, and R. Zebari, "Iot and cloud computing issues, challenges and opportunities: A review," vol. 1, pp. 1–7, 03 2021.
[7] L. Babun, K. Denney, Z. B. Celik, P. McDaniel, and S. Uluagac, "A survey on iot platforms: Communication, security, and privacy perspectives," Computer Networks, vol. 192, p. 108040, 03 2021.
[8] IRENA, "Internet of things innovation landscape brief," International Renewable Energy Agency, Abu Dhabi, 2019.
[9] M. Atiquzzaman, M. Noura, and M. Gaedke, "Interoperability in internet of things: Taxonomies and open challenges," Mobile Networks and Applications, pp. 796–809, 07 2018.
[10] O. Alhazmi and K. Aloufi, "Performance analysis of the hybrid mqtt/uma and restful iot security model," Advances in Internet of Things, vol. 11, pp. 26–41, 01 2021.
[11] G. Salazar Chacón, C. Venegas, M. Baca, I. Rodriguez, and L. Marrone, "Open middleware proposal for iot focused on industry 4.0," 2018 IEEE 2nd Colombian Conference on Robotics and Automation (CCRA), pp. 1–6, 11 2018.
[12] S. Nuratch, "Applying the mqtt protocol on embedded system for smart sensors/actuators and iot applications," 2018 15th International Conference on Electrical Engineering/Electronics, Computer, Telecommunications and Information Technology, no. 1, 2018.
[13] M. Iglesias Urkia, A. Orive, and A. Urbieta, "Analysis of coap implementations for industrial internet of things: A survey," Procedia Computer Science, pp. 188–195, 12 2017.
[14] M. Mónica, G.-R. Carlos, and C. Celeste, "Performance evaluation of coap and mqtt-sn in an iot environment," *Proceedings*, vol. 31, p. 49, 11 2019.
[15] S. Jaikar and R. Kamatchi, "A survey of messaging protocols for iot systems," International Journal of Advanced in Management, Technology and Engineering Sciences, 02 2018.
[16] J. Dizdarevic, F. Carpio, A. Jukan, and X. Masip, "A survey of communication protocols for internet of things and related challenges of fog and cloud computing integration," ACM Computing Surveys, 04 2018.
[17] T. Sultana, "Choice of application layer protocols for next generation video surveillance using internet of video things," IEEE Access, pp. 41607 – 41624, 2019.
[18] Z. B. Babovic, J. Protic, and V. Milutinovic, "Web performance evaluation for internet of things applications," IEEE Access, 2016.
[19] T. Perumal, S. K. Datta, and C. Bonnet, "Iot device management framework for smart home scenarios," pp. 54–55, 2015.
[20] J. de Carvalho Silva, J. Rodrigues, J. Al-Muhtadi, R. Rabelo, and V. Furtado, "Management platforms and protocols for internet of things: A survey," Sensors, vol. 19, p. 676, 02 2019.
[21] M. Aboubakar, M. Kellil, and P. Roux, "A review of iot network management: Current status and perspectives," Journal of King Saud University - Computer and Information Sciences, 2021.
[22] A. Ahmed, "Secured service discovery technique in iot," Journal of Communications, vol. 14, pp. 40–46, 01 2019.
[23] N. Naik and P. Jenkins, "Web protocols and challenges of web latency in the web of things," 2016 Eighth International Conference on Ubiquitous and Future Networks (ICUFN), pp. 845–850, Aug 2016.
[24] N.Naik, "Choice of effective messaging protocols for iot systems: Mqtt, coap, amqp and http," 2017 IEEE International Systems Engineering Symposium(ISSE), pp. 1–7, 2017.
[25] e. a. Esfahani, Alireza, "A lightweight authentication mechanism for m2m communications in industrial iot environment," IEEE Internet of Things Journal, pp. 1–1, 08 2017.
[26] D. Soni and A. Makwana, "A survey on mqtt: A protocol of internet of things(iot)," International Conference on Telecommunication, Power Analysis And Computing Technique (ICTPACT), 2017.
[27] T. Salman and R. Jain, "A survey of protocols and standards for internet of things," 02 2019.
[28] B. Çorak, F. Okay, M. Güzel, S. Murt, and S. Ozdemir, "Comparative analysis of iot communication protocols," pp. 1–6, 06 2018.
[29] T. Prantl, L. Iffländer, S. Herrnleben, S. Engel, S. Kounev, and C. Krupitzer, "Performance impact analysis of securing mqtt using tls," pp. 241–248, 04 2021.
[30] A. Al-Fuqaha, M. Guizani, M. Mohammadi, M. Aledhari, and M. Ayyash, "Internet of things: A survey on enabling technologies, protocols, and applications," IEEE Communications Surveys Tutorials, no. 4, pp. 2347–2376, 2015.
[31] A. Hornsby and R. Walsh, "From instant messaging to cloud computing, an xmpp review," Proceedings of the International Symposium on Consumer Electronics, ISCE, pp. 1 – 6, 07 2010.
[32] Z. B. Babovic, J. Protic, and V. Milutinovic, "Web performance evaluation for internet of things applications," IEEE Access, pp. 6974–6992, 2016.
[33] J. Wytrebowicz, K. Cabaj, and J. Krawiec, "Messaging protocols for iot systems—a pragmatic comparison," Sensors, vol. 21, 10 2021.
[34] E. Al-Masri, K. R. Kalyanam, J. Batts, J. Kim, S. Singh, T. Vo, and C. Yan, "Investigating messaging protocols for the internet of things (iot)," IEEE Access, vol. 8, pp. 94880–94911, 2020.
[35] B. Cabé, "Key trends from the iot developer survey 2018," Available on https://blog.benjamin-cabe.com/2018/04/17/key-trends-iot-developer-survey-2018.
[36] S. Raffaele, "Ibm bluemix the cloud platform for creating and delivering applications," International Technical Support Organization, vol. 6, 08 2015.
[37] T. Pflanzner and A. Kertesz, "A survey of iot cloud providers," 2016 39th International Convention on Information and Communication Technology, Electronics and Microelectronics (MIPRO), pp. 730–735, 2016.
[38] P. Pierleoni, R. Concetti, A. Belli, and L. Palma, "Amazon, google and microsoft solutions for iot: Architectures and a performance comparison," *IEEE Access*, vol. 8, pp. 5455–5470, 2020.
[39] B. Nakhuva and T. Champaneria, "Study of various internet of things platforms," International Journal of Computer Science Engineering Survey, vol. 6, pp. 61–74, 12 2015.
[40] M. Zdravković, M. Trajanovic, J. Sarraipa, R. Jardim-Gonçalves, M. Lezoche, A. Aubry, and H. Panetto *6th International Conference on Information Society and Technology (ICIST 2016)*, 02 2016.

[41] P. Ganguly, "Selecting the right iot cloud platform," *2016 International Conference on Internet of Things and Applications (IOTA)*, pp. 316–320, 2016.

[42] P. Ray, "A survey on internet of things architectures," EAI Endorsed Transactions on Internet of Things, vol. 2, p. 151714, 12 2016.

[43] S. Mohan and K. B R, "A survey on iot platforms," international journal of scientific research and mordern education, vol. 1, 05 2016.

[44] J. Zhou, T. Leppänen, E. Harjula, M. Ylianttila, T. Ojala, C. Yu, H. Jin, and L. Yang, "Cloudthings: A common architecture for integrating the internet of things with cloud computing," Proceedings of the 2013 IEEE 17th International Conference on Computer Supported Cooperative Work in Design, CSCWD 2013, pp. 651–657, 06 2013.

[45] J. Mineraud, O. Mazhelis, X. Su, and S. Tarkoma, "A gap analysis of internet-of-things platforms," Computer Communications, pp. 5–16, 2016.

[46] P. Kostelnik, M. Sarnovsky, and K. Furdík, "The semantic middleware for networked embedded systems applied in the internet of things and services domain," Scalable Computing: Practice and Experience, vol. 12, 09 2011.

[47] "Google iot core," Available on https://cloud.google.com/iot-core, 06 2022.

[48] "Oracle iot," Available on https://www.oracle.com/internet-of-things/, 06 2022.

[49] "Cisco kinetic," Available on https://developer.cisco.com/site/kinetic/, 06 2022.

[50] C. A. L. Putera and F. J. Lin, "Incorporating oma lightweight m2m protocol in iot/m2m standard architecture," 2015 IEEE 2nd World Forum on Internet of Things (WF-IoT), pp. 559–564, 2015.

[51] OMA, "Oma device management protocol approved version 1.2.1," 06 2008.

[52] OMA, "Oma device management protocol approved version 1.3," 05 2016.

[53] OMA, "Device management architecture candidate version 1.3," 03 2012.

[54] O. M. Alliance, "Lightweight machine to machine technical specification," 02 2018.

[55] G. Association, "Iot standard diagnostic logging version 1.0," officiel document," 01 2018.

[56] F. Broadband, "Tr-069 cpe wan management protocol," 03 2018.

[57] F. Broadband, "Tr-069 cpe wan management protocol v1.1," 09 2007.

[58] G. Nebbione and M. C. Calzarossa, "Security of iot application layer protocols: Challenges and findings," Future Internet, vol. 12, no. 3, 2020.

[59] C. Cabrera, A. Palade, and S. Clarke, "An evaluation of service discovery protocols in the internet of things," Proceedings of the Symposium on Applied Computing, p. 469–476, 2017.

[60] A. Bröring, S. K. Datta, and C. Bonnet, "A categorization of discovery technologies for the internet of things," IoT'16: Proceedings of the 6th International Conference on the Internet of Things, p. 131–139, 2016.

[61] D. Kim, H. Lee, and D. Kim, "Enhanced industrial message protocol for real-time iot platform," 2018 International Conference on Electronics, Information, and Communication (ICEIC), pp. 1–2, 2018.

[62] M. Finochietto, G. M. Eggly, R. Santos, J. Orozco, S. F. Ochoa, and R. Meseguer, "A role-based software architecture to support mobile service computing in iot scenarios," Sensors, vol. 19, no. 21, 2019.

[63] B. Konieczek, M. Rethfeldt, F. Golatowski, and D. Timmermann, "Real-time communication for the internet of things using jcoap," 2015 IEEE 18th International Symposium on Real-Time Distributed Computing, pp. 134–141, 2015.

[64] A. Depari, D. Fernandes Carvalho, P. Bellagente, P. Ferrari, E. Sisinni, A. Flammini, and A. Padovani, "An iot based architecture for enhancing the effectiveness of prototype medical instruments applied to neurodegenerative disease diagnosis," Sensors, vol. 19, no. 7, 2019.

[65] U. Hunkeler, H. L. Truong, and A. Stanford-Clark, "Mqtt-s — a publish/subscribe protocol for wireless sensor networks," 2008 3rd International Conference on Communication Systems Software and Middleware and Workshops (COMSWARE '08), pp. 791–798, 2008.

[66] O. Akintade, T. Yesufu, and L. Kehinde, "Development of an mqtt-based iot architecture for energy-efficient and low-cost applications," International Journal of Internet of Things, pp. 27–35, 06 2019.

[67] A. Alshantout and L. Al-Awami, "Enhancing mqtt-sn performance via fountain codes in extreme conditions," 2019 15th International Wireless Communications Mobile Computing Conference (IWCMC), pp. 1184–1189, 2019.

[68] B. Schütz, J. Bauer, and N. Aschenbruck, "Improving energy efficiency of mqtt-sn in lossy environments using seed-based network coding," 2017 IEEE 42nd Conference on Local Computer Networks (LCN), pp. 286–293, 2017.

[69] W. Mardini, M. Bani Yassein, M. N. Alrashdan, A. Alsmadi, and A. Bani Amer, "Application-based power saving approach for iot coap protocol," Proceedings of the First International Conference on Data Science, E-Learning and Information Systems, 10 2018.

[70] R. Giambona, A. Redondi, and M. Cesana, "Mqtt+: Enhanced syntax and broker functionalities for data filtering, processing and aggregation," *Q2SWinet'18: Proceedings of the 14th ACM International Symposium on QoS and Security for Wireless and Mobile Networks*, p. 77–84, 10 2018.

[71] M. A. A. da Cruz, J. J. P. C. Rodrigues, E. S. Paradello, P. Lorenz, P. Solic, and V. H. C. Albuquerque, "A proposal for bridging the message queuing telemetry transport protocol to http on iot solutions," 2018 3rd International Conference on Smart and Sustainable Technologies (SpliTech), pp. 1–5, 2018.

[72] S. Alexandre, F. Carlier, and V. Renault, "Dynamic bridge generation for iot data exchange via the mqtt protocol," Procedia Computer Science, pp. 90–97, 01 2018.

[73] M. Al-Osta, B. Ahmed, and A. Gherbi, "A lightweight semantic web-based approach for data annotation on iot gateways," Procedia Computer Science, pp. 186–193, 12 2017.

[74] S. Wagle, "Semantic data extraction over mqtt for iotcentric wireless sensor networks," 2016 International Conference on Internet of Things and Applications (IOTA), pp. 227–232, 2016.

[75] O. Sadio, I. Ngom, and C. Lishou, "Lightweight security scheme for mqtt/ mqtt-sn protocol," 2019 Sixth International Conference on Internet of Things: Systems, Management and Security (IOTSMS), pp. 119–123, 2019.

[76] C. Fetzer, G. Pfeifer, and T. Jim, "Enhancing dns security using the ssl trust infrastructure," 10th IEEE International Workshop on Object-Oriented Real-Time Dependable Systems, pp. 21– 27, 03 2005.

[77] M. Abubakar, Z. Jaroucheh, A. Al-Dubai, and X. Liu, "Blockchain-based identity and authentication scheme for mqtt protocol," pp. 73–81, 03 2021.

[78] F. Buccafurri, V. De Angelis, and R. Nardone, "Securing mqtt by blockchain-based otp authentication," Sensors, vol. 20, no. 7, 2020.

[79] F. Buccafurri and C. Romolo, "A blockchain-based otp-authentication scheme for constrainded iot devices using mqtt," pp. 1–5, 09 2019.

[80] A. Mektoubi, H. L. Hassani, H. Belhadaoui, M. Rifi, and A. Zakari, "New approach for securing communication over mqtt protocol a comparaison between rsa and elliptic curve," 2016 International Conference on Systems of Collaboration (SysCo), pp. 1–6, 2016.

[81] S. Jakeer Hussain and G. Ramyasri, "Security enhancement for devices using restapi, middleware iot gateway," International Journal of Engineering and Advanced Technology (IJEAT), p. 2249 – 8958, 09 2019.

[82] F. Seidel, K.-F. Krentz, and C. Meinel, "Deep en-route filtering of constrained application protocol (coap) messages on 6lowpan border routers," *2019 IEEE 5th World Forum on Internet of Things (WF-IoT)*, pp. 201–206, 2019.

[83] A. S. Sadeq, R. Hassan, S. S. Al-rawi, A. M. Jubair, and A. H. M. Aman, "A qos approach for internet of things (iot) environment using mqtt protocol," 2019 International Conference on Cybersecurity (ICoCSec), pp. 59–63, 2019.

[84] A. Sadeq, R. Hassan, and A. Jubair, "Enhanced mqtt for providing qos in internet of things (iot): A study," Advanced Science Letters, pp. 5199–5203, 07 2019.

[85] H. Hwang, J. Park, B. Lee, and J. Shon, "An enhanced reliable message transmission system based on mqtt protocol in iot environment," Lecture Notes in Electrical Engineering, pp. 982–987, 01 2017.

[86] K. Hwang, J. M. Lee, I. H. Jung, and D. Lee, "Modification of mosquitto broker for delivery of urgent mqtt message," 2019 IEEE Eurasia Conference on IOT, Communication and Engineering (ECICE), pp. 166–167, 2019.

[87] K. Hwang, J. Moon Lee, and I. Hwan Jung, "Extension of reception frequency control to mqtt protocol," International Journal of Innovative Technology and Exploring Engineering (IJITEE), pp. 2278–3075, 06 2019.

[88] J. Luzuriaga, M. Perez, P. Boronat, J.-C. Cano, C. Calafate, and P. Manzoni, "Improving mqtt data delivery in mobile scenarios: Results from a realistic testbed," Mobile Information Systems, pp. 1–11, 07 2016.

[89]  M. Swarna and T. Godhavari, "Enhancement of coap based congestion control in iot network - a novel approach," Materials Today: Proceedings, 2020.
[90]  A. Oak and R. D. Daruwala, "Assessment of message queue telemetry and transport (mqtt) protocol with symmetric encryption," 2018 First International Conference on Secure Cyber Computing and Communication (ICSCCC), pp. 5–8, 2018.
[91]  D. Swessi and H. Idoudi, "A survey on internet-of-things security: Threats and emerging countermeasures," Wireless Personal Communications, vol. 124, pp. 1–36, 05 2022.
[92]  R. Fotohi, Somayyeh, F. Bari, and M. Yusefi, "Securing wireless sensor networks against denial-of-sleep attacks using rsa cryptography algorithm and interlock protocol," International Journal of Communication Systems, 03 2020.
[93]  T. Vadluri, "A research on interoperability issues in internet of things at application layer," Mobile Netw Appl, 03 2020.